\documentclass[aps, pra, twocolumn, amsmath, graphicx, latexsym, asmsymb, amsfonts]{revtex4}

\usepackage{epsfig}

\begin{document}


\title{Capacities of noiseless quantum channels for massive indistinguishable particles: Bosons vs. fermions}

\author{Aditi Sen(De)$^{1,2}$, 
Ujjwal Sen$^{1,2}$, 
 Bartosz Gromek\(^3\), Dagmar Bru{\ss}$^4$
and Maciej Lewenstein$^{2,5}$}

\affiliation{\(^1\)ICFO-Institut de Ci\`encies Fot\`oniques, 
Mediterranean Technology Park, E-08860  Castelldefels (Barcelona), Spain\\
$^2$Institut f\"ur Theoretische Physik, 
Universit\"at Hannover, D-30167 Hannover,
Germany\\
\(^3\)Department of Theoretical Physics, University of Lodz, ul. Pomorska 149/153, PL-90236  Lodz, Poland\\
$^4$Institut f\"ur Theoretische Physik III, Heinrich-Heine-Universit\"at D\"usseldorf, 
D-40225 D\"usseldorf, Germany\\
\(^5\)ICREA and ICFO-Institut de Ci\`encies Fot\`oniques, 
Mediterranean Technology Park, E-08860  Castelldefels (Barcelona), Spain
}

\begin{abstract}

We consider  information transmission through a noiseless quantum channel,
where the  information is encoded
into massive indistinguishable particles: bosons or fermions. 
We study the situation in which the particles
are noninteracting. 
The encoding input states obey a set of physically motivated constraints on the mean values of the energy and particle number. 
In such a case, the determination of both 
classical and quantum capacity 
reduces 
to a constrained maximization 
of   entropy.
In the case of noninteracting bosons, signatures of 
Bose Einstein  condensation can be observed in the behavior of the capacity. 
A major motivation for these considerations is to compare the 
information carrying capacities of channels that carry bosons
with those that carry fermions. 
We show analytically that fermions generally provide higher channel capacity, i.e., they are better suited for transferring 
bits as well as qubits, in comparison to bosons. This holds 
for a large range of power law potentials, and for moderate to high temperatures.  
Numerical simulations seem to indicate that the result holds for all temperatures.
Also, we consider the low temperature behavior for the three-dimensional box and harmonic trap, and again we show that
the fermionic capacity is higher than the bosonic one  for sufficiently low temperatures.

\end{abstract}
\maketitle

\section{Introduction}

The capacity of  a transmission 
channel is an important quantity in several aspects, ranging from fundamental theoretical 
issues to technology; it is therefore intensively studied in classical \cite{Chennai} and quantum \cite{NC} information theory. 
The capacity of a channel depends 
on different factors, like the type of information that one wants to transmit, the character of the physical 
material that realizes the transfer, the constraints that exist on the channel, the assistance 
that is available for the information transfer, etc.

The capacity of a channel 
for the transmission of classical information
by using quantum states is an important example, which has attracted 
a lot of interest recently. In this respect, a fundamental result, obtained 
more than 30 years ago, is the ``Holevo bound'' \cite{Holevo, Holevo73} 
(see also \cite{onno, Schumacher-er-ghyama, byapari-alada}).
An essential message conveyed by the Holevo bound is that \emph{at most \(n\) bits (binary digits) of 
classical information can be carried by a quantum system of \(n\) distinguishable qubits 
(two-dimensional quantum systems)}.
However, the Holevo bound  includes arbitrary encoding and decoding strategies.
This has the consequence that for information transmission with infinite dimensional systems through noiseless channels, e.g. with the modes 
of an electromagnetic field, the Holevo bound predicts infinite capacities.

A similar situation appears also for the capacity of transmitting quantum information, merely by definition.

To avoid such nonphysical values, it is important to introduce the relevant 
physical constraints, while maximizing over the classical or quantum information that can be encoded in
a physical system. Classical capacities of channels that carry photons in the form of modes
of an electromagnetic field 
have
 been studied extensively in the last decade \cite{onno, rmp, onno2}. 
In this case, as well as for the quantum capacity, as considered in this paper, the natural choice is an energy constraint 
on the input states of the channel.


In this paper, we will deal with channels carrying particles that have 
a non-vanishing mass. Such a problem is motivated by recent 
experiments producing atomic waveguides in optical microstructures \cite{hannover}, or in atom chips \cite{revchip}. These
experiments indicate a new possibility of using massive particles in quantum channels for classical as well as quantum communication over
macroscopic, or at least mesoscopic distances. 
For such massive particles,  it is natural to put  an average particle number constraint along with an average energy constraint
on the input state of the channel.
We will show that with these constraints, the grand canonical ensemble of statistical mechanics attains the classical as well as quantum 
capacity of 
noiseless channels
that carry trapped noninteracting massive particles.
Moreover, we show that the classical and quantum capacities in the case of noiseless quantum channels (with the above physical constraints) are 
the same.
Therefore in the remainder of this paper, the phrase ``capacity of (quantum) channels carrying 
massive particles'' (usually implicitly implying that the constraints on energy and particle numbers are already taken into account),
unless with the specific adjective,
will mean the classical as well the quantum capacity.
We should mention here that sometimes, it is more natural to consider the average energy constraint with a fixed number of particles. 
In this case, the capacity is attained in the canonical ensemble. Calculations of different 
properties of the canonical ensemble are
usually very hard, as compared to the grand canonical ensemble.
However, several studies indicate  
that the two ensembles give similar values for the average occupation numbers even for finite, but moderately large number of particles
\cite{Nobel-kajer-por}.
However, there are physical quantities of interest that give drastically different values. 
For example, the fluctuation in the average number of particles in a Bose Einstein condensate is unphysically large 
according to the grand canonical ensemble, while that in the canonical ensemble is physically meaningful \cite{Nobel-kajer-por}. 
In this paper, we will restrict ourselves to the
grand canonical ensembles, for finding the capacities of channels carrying massive particles. 
The capacities, being dependent only on the average occupation  numbers, will be similar already for a moderately large 
number of particles.

Noninteracting bosons exhibit a critical phenomenon, the Bose Einstein condensation (BEC). 
In this paper, we will show that the capacity of channels carrying 
noninteracting bosons also exhibit this criticality with temperature. The capacity
changes its behavior from being 
concave to convex (with respect to temperature), at the critical temperature. Noninteracting  fermions do not
show any critical behavior. However, interacting fermions can exhibit the Bardeen-Cooper-Schrieffer (BCS) 
transition. As we have shown in Ref. \cite{amader}, the channel capacity in this case also indicates the BCS transition.
It should be noted here that the convexity and concavity, that we speak about in this paper, are 
always with respect to temperature, and not with respect to mixing of states.

An important motivation of this study is  to compare the capacities of channels carrying 
bosons with that of fermions. We will obtain the following result analytically: 
\\
\noindent \emph{Capacities of noiseless quantum channels carrying noninteracting spinless fermions 
are higher than that of spinless bosons for a wide range of power law potentials 
 for sufficienly high temperatures.
}

Since we  obtain the above result by using perturbation theory up to the third order, it holds already for moderate temperatures. 
Note  that our numerical simulations indicate that fermions are better carriers of information than  
bosons even in the low temerature region.

It may be noted that although the quantum channel that we consider in the paper is noiseless, 
the states that are used to encode the information are allowed to be noisy (mixed states encoding).
In the case of classical capacity, it will turn out that there exist ensembles of pure states, which can be used to encode the 
classical information (to be 
transferred), for attaining the maximal information transfer.

The paper is organized as follows. In Sec. \ref{sec-Holevo}, we discuss the Holevo bound on classical information transmission. 
We define the 
classical capacity of quantum channels in the next section (Sec. \ref{sec-jomey-doi}). In Sec. \ref{quantumcapa},
we define quantum capacity of quantum channels.
In Sec. \ref{sec-fundamental}, we  discuss the important 
 ensembles in statistical mechanics, and in its first three subsections, we discuss respectively the microcanonical, the canonical, and the 
grand canonical ensembles. In Sec. \ref{sub-ghuri}, we consider the situation when there is a constraint on the  average  number of 
particles, while there is \emph{no} constraint on the (exact or average) number of particles.
In the next section (Sec. \ref{kapakapa}), we discuss the capacities of quantum channels with constrained inputs.
We then briefly discuss the capacities of channels 
 carrying photons in Sec. \ref{sec-photo}. In Sec. \ref{sec-mota}, we 
 consider  spinless noninteracting bosons and fermions, and calculate the capacity of channels in these cases. In Sec. \ref{sec-ghyama},
we prove a theorem stating that fermions can carry more information (classical and quantum) than bosons. 
In Sec. \ref{sec-jotota-bhebechhilum-totota-soja-noi}, we consider the low temperature case, for the 3D box and the 3D harmonic trap. For both 
these cases, the fermionic capacity is again higher than the bosonic one for sufficiently low temperatures.
In the last section 
(Sec. \ref{sec-sesh}), we make  some concluding remarks.
This paper presents on one hand the results of Ref. \cite{amader} with more details, and on the other 
hand generalizes them to the case of trapped 
non-interacting bosonic and fermionic gases for a large class of power law potentials. Let us note that many of 
such potentials are feasible with currently available technology. We will see that the signature of BEC in the channel capacity 
in the case of a harmonic 
trap, is more pronounced 
than that in the case of a uniform trap, as 
considered in Ref. \cite{amader}.

It may be worthwhile to mention here that the mathematics required for obtaining the capacities,
is similar to that in quantum statistical mechanics, where one maximizes the entropy under different constraints. 
This is not very surprising, as the bounds on information transfer, discussed in Secs. \ref{sec-Holevo} and \ref{quantumcapa}, 
are ``entropy-like'' quantities.


\section{The Holevo bound}
\label{sec-Holevo}

The Holevo bound is an upper bound on the amount of classical information that can be accessed from a quantum ensemble in which the information is 
encoded. 
Suppose therefore that a sender Alice (\(A\)) obtains the classical message \(i\), and she knows that 
this happens with probability \(p_i\). 
She wants to send it to a receiver Bob (\(B\)). To do so, Alice encodes the information \(i\) in a quantum state \(\rho_i\), and sends 
the quantum state to Bob. 
Bob receives the ensemble \(\{p_i, \rho_i\}\), and wants to obtain as much information as possible about \(i\), for which 
he performs a measurement, that gives the result \(m\), with probability \(q_m\). Let the corresponding post-measurement ensemble be 
\(\{p_{i|m}, \rho_{i|m}\}\). The classical information gathered can be quantified by the mutual information between the 
message index \(i\) and the measurement outcome \cite{Chennai}:
\begin{equation}
I(i:m)= H(\{p_i\}) - \sum_m q_m H(\{p_{i|m}\}).
\end{equation}
Here \(H(\{r_x\}) = -\sum_xr_x\log_2r_x\) is the Shannon entropy of the probability distribution \(\{r_x\}\).
Throughout the paper, we calculate the all the quantities on amounts of information transfer in bits (binary digits).
Note that the mutual information can be seen as the difference between the initial disorder and the (average) final disorder. 
Bob will be interested to obtain the maximal information, which is the maximum of \(I(i:m)\) over all measurement strategies. This 
quantity is called the accessible information:
\begin{equation}
I_{acc} = \max I(i:m),
\end{equation} 
where the maximization is over all measurement strategies.

The maximization involved in the definition of accessible information is usually hard to compute, and hence it is important to know bounds 
on \(I_{acc}\) \cite{Holevo, Holevo73, Utpakhi}. 
In particular, in Ref. \cite{Holevo, Holevo73}, a universal upper bound, called the Holevo bound (or Holevo quantity), on \(I_{acc}\) is 
given (see also \cite{onno, Schumacher-er-ghyama, byapari-alada}): 
\begin{equation}
I_{acc}(\{p_i, \rho_i\}) \leq \chi(\{p_i, \rho_i\}) \equiv S(\overline{\rho}) - \sum_i p_i S(\rho_i).
\end{equation}
Here \(\overline{\rho} = \sum_ip_i\rho_i\) is the average ensemble state, and 
\begin{equation}
\label{amar-byala-je-jai}
S(\varsigma)= - \mbox{tr}(\varsigma \log_2 \varsigma)
\end{equation} 
is the von Neumann entropy of \(\varsigma\).


\section{Classical capacity of a quantum channel: Unconstrained inputs}
\label{sec-jomey-doi}

Consider a  quantum channel \({\cal R}\) that acts on  \(d\)-dimensional quantum systems as 
inputs. Suppose that Alice wants to send some 
classical information \(i\), that occurs with probability \(p_i\), through this quantum channel to Bob. She encodes 
this classical information in the quantum state \(\rho_i\), where the Hilbert space corresponding to the quantum states \(\rho_i\)
is \(d\)-dimensional. The classical capacity of this quantum channel is the maximal classical information that can be sent through this channel, and 
is therefore the accessible information of the ensemble \(\{p_i, {\cal R} (\rho_i)\}\), 
maximized over all such ensembles on the \(d\)-dimensional Hilbert 
space \({\cal H}^d\):
\begin{equation}
\label{kachuri1_hanuman_haluikar}
{\cal C} = \max_{\{p_i,\rho_i\} \phantom{o} \mbox{\footnotesize{on}} \phantom{o}  {\cal H}^d} I_{acc} (\{p_i, {\cal R} (\rho_i)\}).
\end{equation}
However, capacities are usually defined in an asymptotic sense. Therefore, 
the classical capacity in this case is
\begin{equation}
\label{kachuri1}
{\cal C}^{\infty} = \lim_{n \to \infty}
\phantom{p} \frac{1}{n} \phantom{p}
\max_{\{p_i,\rho_i\} \phantom{o} \mbox{\footnotesize{on}} \phantom{o} \left( {\cal H}^d \right)^{\otimes n} } I_{acc} (\{p_i, {\cal R} (\rho_i)\}).
\end{equation}

The Holevo bound implies that  
\begin{eqnarray}
\label{kachuri_moronodolai}
{\cal C} & \leq & \max_{\{p_i,\rho_i\} \phantom{o} \mbox{\footnotesize{on}} \phantom{o}  {\cal H}^d} 
\left[S({\cal R}(\overline{\rho})) - \sum_i p_iS( {\cal R}(\rho_i)) \right],  
\end{eqnarray}
and
\begin{eqnarray}
\label{kachuri_dhorey_rashigachhi}
&& {\cal C}^{\infty}  \leq  \nonumber \\
&& \lim_{n \to \infty}
\phantom{p} \frac{1}{n} \phantom{p}
\max_{\{p_i,\rho_i\} \phantom{o} \mbox{\footnotesize{on}} \phantom{o} \left( {\cal H}^d \right)^{\otimes n} }
%
%
\left[S({\cal R}(\overline{\rho})) - \sum_i p_iS( {\cal R}(\rho_i)) \right]. \nonumber \\
&&
\end{eqnarray}

The quantity \({\cal C}^{\infty}\) is usually very difficult to handle. 
The Holevo-Schumacher-Westmoreland theorem \cite{babarey, maarey} states that in the particular case when the inputs are products on the tensor product Hilbert space
\(\left( {\cal H}^d \right)^{\otimes n}\), the capacity (let us denote the capacity in this case by \({\cal C}_d^{\infty, 1}\))
is given by 
\begin{equation}
\label{kachuri1_benaras}
{\cal C}^{\infty, 1} = 
\max_{\{p_i,\rho_i\} \phantom{o} \mbox{\footnotesize{on}} \phantom{o}  {\cal H}^d} 
\left[S({\cal R}(\overline{\rho})) - \sum_i p_iS( {\cal R}(\rho_i)) \right].
\end{equation}

For the case of the noiseless quantum channel that carries \(d\) dimensional quantum states noiselessly, 
all the above capacities equal to \(\log_2 d\). This value  
is attained by any complete orthogonal basis of pure states on 
\({\cal H}^d\).


Now for infinite dimensional quantum systems, the channel capacity obtained in this way predicts an unphysical infinite capacity. 
This is because the Holevo bound itself does not include any constraint on the available physical resources in an actual 
implementation of the information transfer. In particular, arbitrary encoding and decoding schemes are allowed. 
To avoid this infinite capacity, one usually maximizes the accessible information over all ensembles that satisfies certain 
physical constraints. Due to the form of the Holevo bound on accessible information, such constrained maximizations are 
very similar to the ones in statistical mechanics. The same is true for the case of the quantum capacity
 which we briefly discuss in the succeeding section, and then in Sec. \ref{sec-fundamental}, we 
 briefly discuss some  similar constrained maximizations of statistical mechanics.

\section{Quantum Capacity of a quantum channel: Unconstrained inputs}
\label{quantumcapa}

We now consider the case of sending qubits (as opposed to bits) using quantum channels. 
The quantum capacity  
can be considered in (at least) the following  four different situations \cite{Horodecki_private}: 
the quantum channel \({\cal R}\) (acting on \(d\) dimensional quantum states at its input)  
\begin{itemize}
\item without the help of additional classical communication (in this case, we call the quantum capacity  $Q^{0}$),
\item with an arbitrary amount of forward classical communication ($Q^{\rightarrow}$), 
\item with an arbitrary amount of backward classical communication ($Q^{\leftarrow}$), 
\item with an arbitrary amount of both-way classical communication ($Q^{\leftrightarrow}$). 
\end{itemize}
Let us define the first case, the other definitions being similar.
So, the quantum capacity \(Q^{0}\) is \cite{Shor_lecture}
\begin{equation}
\label{QCformula}
  Q^{0} = \sup \lim_{n \rightarrow \infty}  \frac{\log_2 D(n)}{n},
\end{equation}
where the supremum is over all such cases when there exists a 
$D(n)$ dimensional subspace $\mathbb{S}(n)$, of the total input space  $(\mathcal{H}^d)^{\otimes n}$, 
satisfying the average fidelity criterion
\begin{equation}
\label{Shor_bhodrolok_chhilen_bhag-gish}
\lim_{n \to \infty} 
\int_{|\psi\rangle \in \mathbb{S}(n)} 
\langle \psi | \mathcal{R}^{\otimes n}(|\psi \rangle \langle \psi |) | \psi \rangle d|\psi\rangle =1,
\end{equation}
where no classical communication was used in transferring the input state from the sender to the receiver.

In case of a noiseless channel, the fidelity criterion is automatically satisfied, and all the quantum 
capacities are equal to \(\log_2 d\). Now, if the dimension $d$ of the input space 
is infinite, the capacities are again infinite, as in the case of classical capacity.

There are several remarkable results that are known in the noisy case. In particular, 
\(Q^{0}\) is the maximum coherent information
\cite{Lloyd,Shor_lecture,Devatak03, BarnumNielsen}:
\begin{equation}
\label{dusho-dosh-lakh}
Q^{0} = \max_\rho \left[S(\mathcal{R}(\rho)) - S(\mathcal{I}\otimes \mathcal{R}(\Phi_\rho))\right],
\end{equation}
where the maximization is over all quantum states \(\rho\) defined on the Hilbert space \({\cal H}^d\), 
\(\Phi_\rho\) is a purification of $\rho$, and \(\mathcal{I}\) is the identity operator acting on quantum states
of the ancillary Hilbert space that is required for the purification.
Furthermore
\cite{gachhey-tuley-moi-kerdey-nao, amar-sontan-jyano-thhakey-dhudhey-bhatey},
\begin{equation}
\label{dusho-dosh-lakh-sishu}
Q^{0} = Q^{\rightarrow}.
\end{equation}
%

\section{The fundamental ensembles in statistical mechanics}
\label{sec-fundamental}

In this section, we discuss  the fundamental ensembles in statistical mechanics (see e.g. 
\cite{Huang, molla-nasiruddin, Reichl}). 

\subsection{The microcanonical ensemble}

Suppose that we have a physical system described by the Hamiltonian \({\cal H}\). We assume that the system has a fixed 
particle number \(N\), and a fixed energy \(E\). (To be more general,
one must allow for fluctuations around \(E\) and \(N\).)
 This represents a closed and isolated system. We want to find the state \(\varrho_{MC}\) of 
the system such that it maximizes the von Neumann entropy \(S\). Let \(|E,N,k\rangle\) denote the state a energy \(E\) 
and particle number \(N\), and where \(k\) enumerates the degeneracy. 
Let \(\Omega(E,N)\) be the total number of orthogonal states with energy \(E\) and particle number \(N\), 
so that \(k= 1,2, \ldots, \Omega(E,N)\). 
Since the system is isolated (as it has fixed energy and fixed number of particles), 
the set \(\{|E,N,k\rangle\}_{k=1}^{\Omega(E,N)}\) spans the allowed Hilbert space. 
Consequently, the maximum entropy is reached by the state 
\begin{equation}
\varrho_{MC} = \frac{1}{\Omega(E,N)} \sum_{k=1}^{\Omega(E,N)} |E,N,k\rangle \langle E,N,k |,
\end{equation}
which is actually the identity on the allowed Hilbert space.
This is the microcanonical state of the system, and the ensemble \(\{1/\Omega(E,N), |E,N,k\rangle\}_{k=1}^{\Omega(E,N)}\) is called the 
microcanonical ensemble.



\subsection{The canonical ensemble}
\label{sec-gojikot}

Consider next a physical system, described by the Hamiltonian \({\cal H}\), which has a fixed 
particle number \(N\) and a fixed \emph{average} energy \(E\). The average energy constraint forces every state \(\varrho\)
to follow
\begin{equation}
\label{eta-holo-energy}
\mbox{tr}({\cal H} \varrho) =E.
\end{equation}
This physical system represents a closed,  but not isolated system. We again want to find the state \(\varrho_{C}\) of 
the system such that it maximizes the von Neumann entropy \(S\). One finds that \cite{Huang}
\begin{equation}
\varrho_{C} = \frac{1}{{\cal Z}_C}\exp(-\beta {\cal H}),
\end{equation}
the canonical state of the system.
 Here \(\beta = \frac{1}{k_BT}\), with \(k_B\) being the Boltzmann constant, and \(T\) the absolute temperature. 
  The canonical partition function  \({\cal Z}_C\) is  given by 
\begin{equation}
\label{snuopoka}
{\cal Z}_C = \mbox{tr} \left(\exp(-\beta {\cal H})\right).
\end{equation}
The state \(\varrho_C\) is called the canonical state of the system and the ensemble 
\(\{\frac{1}{{\cal Z}_C}\exp(-\beta \epsilon_i), |\epsilon_i\rangle\}\) is called the canonical ensemble, 
where \(\{\epsilon_i, |\epsilon_i\rangle\}\) is the eigensystem of the Hamiltonian \({\cal H}\).
Just like for the microcanonical ensemble,
 the particle number enters the calculations via the Hamiltonian and for determining the allowed Hilbert space.  
For example, in the calculation of the trace in Eq. (\ref{snuopoka}), the summation runs over all combinations of different numbers of particles
at different energy levels, under the constraint that the sum of all particles in all the levels is \(N\). 
Moreover, for indistinguishable particles, we must also take care about the statistics of the particles.   
For example, if we have a trap, whose energy levels are \(\epsilon_i\), and in which \(N\) noninteracting spinless bosons are trapped, so that 
the bosons are described by the Hamiltonian \({\cal H}_{b}= \sum_i \epsilon_i a_i^\dagger a_i\) (where \(a_i^\dagger\) and \(a_i\) are 
creation and destruction operators of the \(i\)th mode), 
we have
\begin{eqnarray}
{\cal Z}_C^b &=& \underbrace{\sum_{n_0=0,1,\ldots} \quad \sum_{n_1=0,1,\ldots} \ldots \quad}_{\sum_i n_i = N}   
\exp\left(- \beta \sum_i \epsilon_i n_i\right) \nonumber \\ \\
   &=&  \underbrace{\sum_{n_0=0,1,\ldots} \quad \sum_{n_1=0,1,\ldots} \ldots \quad}_{\sum_i n_i = N}  \prod_i \exp (-\beta \epsilon_i n_i)  ,
\end{eqnarray}
where \(n_i\) is the number of particles at energy level \(\epsilon_i\). 
For a given value of energy \(E\), the temperature \(T\) 
is given by \(E = - \frac{\partial}{\partial \beta}(\log_e {\cal Z}_C)\).

\subsection{The grand canonical ensemble}
\label{sec-danguli}

The next step is  to consider a physical system, described by the Hamiltonian
\({\cal H}\), which has a fixed 
\emph{average} particle number \(N\) and a fixed \emph{average} energy \(E\). The average energy constraint forces every state \(\varrho\)
to follow Eq. (\ref{eta-holo-energy}), while the average particle number constraint reads 
\begin{equation}
\mbox{tr}({\cal N} \varrho) =N,
\end{equation}
where \({\cal N}\) represents the total particle number operator. For example, for a system of spinless 
bosons, \({\cal N} = \sum_i a_i^\dagger a_i \), 
where \(a^\dagger_i\) and \(a_i\) are the creation and annihilation operators of the \(i\)th mode.
This physical system represents an open system. We again want to find the state \(\varrho_{GC}\) of 
the system such that it maximizes the von Neumann entropy \(S\). One finds that the state is \cite{Huang}
\begin{equation}
\varrho_{GC} = \frac{1}{{\cal Z}_{GC}}\exp(-\beta ({\cal H} - \mu {\cal N})),
\end{equation}
i.e., the grand canonical state of the system. 
Here \(\mu\) is the chemical potential.  The grand canonical partition function  \({\cal Z}_{GC}\) is  given by 
\begin{equation}
{\cal Z}_{GC} = \mbox{tr} \left(\exp(-\beta ({\cal H} - \mu {\cal N}))\right).
\end{equation}
Note that in this case, the number of particles is not fixed, and in particular,
\begin{eqnarray}
{\cal Z}_{GC}^b &=& 
\sum_{n_0^b=0}^{\infty} \sum_{n_1^b=0}^{\infty} \ldots    
\exp\left(-\beta \sum_i\left(\epsilon_i - \mu^b \right)n_i^b\right) \nonumber \\
&=&  \prod_i \frac{1}{1 - e^{-\beta (\epsilon_i - \mu^b)}} ,
\end{eqnarray}
in the case of noninteracting spinless 
bosons in a trap of energy levels \(\{\epsilon_i\}\), and where we have now denoted the chemical potential by 
\(\mu^b\).
The ensemble 
\begin{equation}
\left\{\frac{1}{{\cal Z}^b_{GC}} \exp\left(-\beta \sum_i(\epsilon_i - \mu^b) n_i^b\right), \left|n_0^b, n_1^b, \ldots, n_i^b, \ldots 
\right\rangle\right\}
\end{equation}
is the grand canonical ensemble, where the elements of the ensemble runs over all 
combinations of the \(n_i^b\)'s (\(n_i^b = 0, 1, 2, \ldots, \infty\)).

As opposed to the case of the canonical partition function, the grand canonical partition function requires the value of the 
chemical potential. 
We can determine the chemical potential from the constraint on  the average particle number constraint, for 
given values of the temperature and average particle number. 
We can then find the grand partition function with these values. 
The average occupation numbers, in the case of noninteracting spinless bosons, are given by 
\begin{equation}
\overline{n_i^b} = \frac{1}{e^{\beta(\epsilon_i - \mu^b)} -1}.
\end{equation}
The average energy 
is given  by \(E= \sum_i \overline{n_i^b} \epsilon_i\). Otherwise, for given values of the average energy and average particle number, 
the dual constraints on average energy and particle number can be simultaneously solved to obtain the chemical potential, and  the temperature.

\subsection{The constraint on the number of particles}
\label{sub-ghuri}

One may now consider the case, when we do not have a constraint on the number of particles. In other words, arbitrary numbers of particles 
are allowed. For the energy, we still assume that the average energy is fixed to \(E\). In this case, maximizing the entropy leads to the 
partition function
\begin{eqnarray}
\label{SatyenBose}
{\cal Z}^b &=& \sum_{n_0=0}^{\infty} \sum_{n_1=0}^{\infty} \ldots\, 
\exp\left(-\beta \sum_i\epsilon_i n_i\right) \\
\label{Satyendranath}
 &=&  \prod_i \sum_{n_i =0}^{\infty} \exp \left( - \beta \epsilon_i  n_i \right)   , 
\end{eqnarray}
for noninteracting bosons in a trap of energy levels \(\{\epsilon_i\}\). 
Note that this partition function is different from the canonical partition function in that 
there is no constraint on the number of particles (i.e. the constraint \(\sum_i n_i =N\) does not exist anymore), 
and from the grand canonical partition function  
in that the term
\(\beta \mu {\cal N}\) is absent inside the exponent.

In the case when the lowest energy level is zero or negative, at least one of the sums in Eq. (\ref{Satyendranath}) is divergent. 
Such divergence does not occur in \({\cal Z}_C\), due to the constraint \(\sum_i n_i =N\). Neither does it occur 
in \({\cal Z}_{GC}\), as the term \(\beta \mu {\cal N}\) in the exponent ``cures'' the divergence. 

If the lowest energy level is 
positive, we have 
\begin{equation}
\label{eq-parti-jhamela}
{\cal Z}^b = \prod_i \frac{1}{1 - e^{- \beta \epsilon_i}},
\end{equation}
and the average occupation number of the \(i\)th level is
\begin{equation}
\label{eq-ocuu-jhamela}
\overline{n_i} = - \frac{\partial}{\partial (\beta \epsilon_i)} \log_e {\cal Z}^b = \frac{1}{e^{\beta \epsilon_i} -1},
\end{equation}
so that the total average number of particles is
\begin{equation}
N = \sum_i \overline{n_i} = \sum_i \frac{1}{e^{\beta \epsilon_i} -1}, 
\end{equation}
while the total average energy is 
\begin{equation}
 \sum_i \overline{n_i} \epsilon_i = \frac{\epsilon_i}{e^{\beta \epsilon_i} -1}.
\end{equation}
Since all the energy levels are positive, these sums are convergent, and so we are not able to approach the ``thermodynamic limit''
(\(N \to \infty\)).
The thermodynamic limit, i.e. the limiting case of an infinite number of particles 
can be reached only in the trivial case when \(T \to \infty\). 
This for example is the case for photons, in the form of modes of the  electromagnetic field in a closed (finite) cavity, in which case
the zero frequency mode is not populated.
With increasing cavity size, one may approach as close as possible to the zero frequency mode, but then the density matrix describing 
the system ceases, in the limit, to be trace-class.

This problem does not arise in case of the canonical state, as in that case, if the lowest energy level is nonzero (positive or negative), 
we can define \(\epsilon_i^{0} = \epsilon_i - \epsilon_0\) (\(\epsilon_0\) is the lowest energy level), so that 
we can rewrite the canonical partition function as 
\begin{eqnarray}
\label{Sidho}
{\cal Z}^b_C &=& 
\exp(-\beta \epsilon_0 N) \times \nonumber \\
&&\underbrace{\sum_{n_0=0,1,\ldots} \quad \sum_{n_1=0,1,\ldots} \ldots \quad}_{\sum_i n_i = N}   
\exp\left(- \beta \sum_i \epsilon_i n_i\right)
\nonumber \\
 &=&  \exp(-\beta \epsilon_0 N) {\cal Z}^b_{C_0},
\end{eqnarray}
where \({\cal Z}^b_{C_0}\) is the canonical partition function of a system of \(N\) noninteracting bosons in a trap with energy levels 
\(\{\epsilon_i^{0}\}\), and with the lowest energy level vanishing. The  factor  \(\exp(-\beta \epsilon_0 N)\) in Eq. (\ref{Sidho}) is a 
constant for given particle number and energy. 
For the case of the grand canonical state, if the lowest energy is nonzero, we again rewrite the grand canonical partition function as 
\begin{eqnarray}
\label{Kanho}
{\cal Z}^b_{GC} &=&  \prod_i \sum_{n_i = 0}^{\infty} \exp (-\beta (\epsilon_i^{0} - \mu^b_0) n_i) \nonumber \\
&=&  {\cal Z}^b_{GC_0},
\end{eqnarray}
where \({\cal Z}^b_{GC_0}\) is the grand canonical partition function of a system of \(N\) noninteracting bosons in a trap with energy levels 
\(\{\epsilon_i^{0}\}\), and with the lowest energy level vanishing, and with a correspondingly changed chemical potential \(\mu^b_0\).

The discussion  in this subsection
so far  concerned  the case of noninteracting bosons. In the case of noninteracting fermions, 
the Pauli exclusion principle forces the \(n_i\)'s in Eq. (\ref{Satyendranath}) to run from \(0\) to \(1\), so that 
the partition function in this case is (we consider ``spinless'' i.e. polarized fermions)
\begin{eqnarray}
\label{EnricoFermi}
{\cal Z}^f &=&  \prod_i \sum_{n_i =0}^{1} \exp \left( - \beta \epsilon_i  n_i \right)   \nonumber \\
&=& \prod_i \left(  1 + \exp(- \beta \epsilon_i)  \right).
\end{eqnarray}
The average occupation numbers in this case are (we are using the same notation for bosons and fermions)
\begin{equation}
\overline{n_i} = - \frac{\partial}{\partial (\beta \epsilon_i)} \log_e {\cal Z}^f = \frac{1}{e^{\beta \epsilon_i} + 1},
\end{equation}
so that the total number of particles and the total energy have fixed values (for a given temperature), irrespective of the 
value of the lowest energy level. Therefore, again we are not able to approach the thermodynamic limit, unless \(T \rightarrow \infty\).

\section{Capacities of quantum channels with constrained inputs}
\label{kapakapa}

We now consider the case of quantum channels with constrained inputs. The motivation for 
such consideration is the fact that in the noiseless case, the capacities (both classical and quantum) are in some cases
 infinite, and 
thus calls for putting some realistic constraints on the set-up. 

Surely, the constraints will depend on the specific realization of the physical system under consideration.
In anticipation of the cases that we will consider below, we put constraints on the average energy and the 
average number of particles 
of the input states. Suppose therefore that we have a system of particles that are sent through a channel \({\cal R}\), and that is
governed by the Hamiltonian \(\mathbb{H}\). Let the corresponding total particle number operator be \(\mathbb{N}\).
We only allow those inputs, \(\rho\), to the channel that satisfies the following two constraints: 
\begin{equation}
\label{chorer-mayer-boro-gola}
\mbox{tr} (\mathbb{H}\rho) =E,
\end{equation}
and 
\begin{equation}
\label{gan-dhorechhen-gris(m)okaley}
\mbox{tr} (\mathbb{N}\rho) =N,
\end{equation}
for certain chosen values of \(E\) and \(N\).

Consequently, the classical capacities \({\cal C}\) and \({\cal C}^{\infty}\),   are now 
respectively the same as the right-hand-sides of Eqs. (\ref{kachuri1_hanuman_haluikar}) and (\ref{kachuri1}),  
\emph{but} where the 
maximizations are over those ensembles whose average ensemble states satisfy the above constraints. 
For the case of \({\cal C}\), the input ensembles must satisfy exactly the constraints in 
Eqs. (\ref{chorer-mayer-boro-gola}) and (\ref{chorer-mayer-boro-gola}). For the case of 
\({\cal C}^\infty\), the constraints must be suitably modified. In this case, they are
(cf. Ref. \cite{Holevonewconstraint}) 
\begin{equation}
\label{chorer-mayer-boro-gola_koralobodoni}
\mbox{tr} \left(\sum_{i=1}^{n}\mathbb{H}_i\rho \right) =nE,
\end{equation}
and 
\begin{equation}
\label{gan-dhorechhen-gris(m)okaley_koralobodoni}
\mbox{tr} \left(\sum_{i=1}^{n}\mathbb{N}_i\rho \right) =nN,
\end{equation}
where \(\mathbb{H}_i = I \otimes I \otimes \ldots \mathbb{H} \ldots \otimes I\),
i.e., it is the identity operator \(I\) at all the \(n\) sites, except at the \(i\)th one, where it is \(\mathbb{H}\).
Similarly, 
\(\mathbb{N}_i = I \otimes I \otimes \ldots \mathbb{N} \ldots \otimes I\).

It seems plausible that (cf. Eq. (\ref{kachuri1_benaras})) 
\begin{equation}
\label{tutankhamen-kobekar-lok-chhilo-jyano}
{\cal C}^{\infty, 1} = 
\max 
\left[S({\cal R}(\overline{\rho})) - \sum_i p_iS( {\cal R}(\rho_i)) \right],
\end{equation}
where the maximization is over all ensembles \(\{p_i,\rho_i\}\)
such that the average ensemble state \(\overline{\rho}\) satisfies Eqs. 
(\ref{chorer-mayer-boro-gola}) and (\ref{gan-dhorechhen-gris(m)okaley}).
The case of a noiseless channel with constrained inputs was considered in Ref. \cite{Holevonewconstraint}.
(Note that we are using the same notations for the capacities as those in the unconstrained case.)
In the succeeding sections, we will consider the right-hand-side of Eq. (\ref{tutankhamen-kobekar-lok-chhilo-jyano}) to be 
the ``classical capacity'' of the noiseless quantum channel where the inputs are constrained by either or both 
the energy and number constraints.

Similarly, it seems to be true that the quantum capacities (without, or with forward, 
classical communication) of a quantum channel with inputs constrained by 
Eqs. (\ref{kachuri1_hanuman_haluikar}) and (\ref{kachuri1}),  
are given by the Eqs. (\ref{dusho-dosh-lakh}) and (\ref{dusho-dosh-lakh-sishu}) respectively, 
but where the maximization in the right-hand-side of Eq. (\ref{dusho-dosh-lakh}) is 
over those inputs that satisfies Eqs. (\ref{kachuri1_hanuman_haluikar}) and (\ref{kachuri1}).
It is indeed true in the noiseless channel case, as we will show in Sec. \ref{sec-mutki}. (Actually, in the noiseless case,
all the quantum capacities 
defined in Sec. \ref{quantumcapa} are equal.)

\section{Communication channels that carry photons}
\label{sec-photo}

As we have already discussed in Secs. \ref{sec-jomey-doi} and \ref{quantumcapa}, 
the classical and quantum capacities of a noiseless quantum channel that are constrained 
only by the dimension of the transferred state, predict infinite capacities. To avoid such infinite capacities, in the case of 
communication channels that carry photons, one usually applies the energy constraint on the allowed states (or ensembles) that are fed into the 
quantum channel \cite{onno, rmp, onno2}. Suppose therefore that we have a system of photons that is being sent through a channel, and is 
described by the Hamiltonian
\({\cal H}_p = \sum_i \hbar \omega_i a^\dagger_i a_i\). Here, we consider the case of classical capacity.
From the considerations for massive bosons considered in Sec. \ref{sec-mutki}, it will be apparent that 
all the quantum capacties defined in Sec. \ref{quantumcapa} are equal to the classical capacity, in this case of photons also. 

The energy constraint implies that 
for any ensemble \(\{p_i, \rho_i\}\) used in a classical information transfer must satisfy 
\begin{equation}
\label{urdon-tubri}
\mbox{tr} \left({\cal H}_p \overline{\rho}\right) = E,
\end{equation}
where \(E\) is a pre-assigned value of average energy for the system, and where \(\overline{\rho} = \sum_i p_i \rho_i\) is the 
average ensemble state. The classical 
 capacity in this case is therefore obtained by maximizing the Holevo quantity \(\chi(\{p_i, \rho_i\})\)
over all possible ensembles \(\{p_i, \rho_i\}\) that satisfy the average energy constraint in Eq. (\ref{urdon-tubri}):
\begin{equation}
{\cal C}^p_E = \max_{\{p_i, \rho_i\} \mbox{\footnotesize{ satisfying Eq. (\ref{urdon-tubri}) }}} 
\chi (\{p_i, \rho_i\}).
\end{equation}
Therefore,
\begin{equation}
\label{chulbul}
{\cal C}^p_E \leq \max_{\{p_i, \rho_i\} \mbox{\footnotesize{ satisfying Eq. (\ref{urdon-tubri}) }}}
S(\overline{\rho}).
\end{equation}

However, the maximization on the  rhs of the inequality (\ref{chulbul}) is the same as has already been considered in Sec. 
\ref{sub-ghuri}, since the number of photons is not conserved.
The ensemble that maximizes the rhs of Eq. (\ref{chulbul})  consists of the number states (Fock states), which are 
mutually orthogonal. Therefore, the capacity \({\cal C}^p_E\) is  attained for this ensemble:
\begin{equation}
{\cal C}^p_E = H\left(\left\{ \frac{1}{{\cal Z}_b}\exp\left(- \sum_i \beta\hbar\omega_i n_i \right) \right\}\right).
\end{equation} 
Here 
\({\cal Z}^b\) is given by  Eq. (\ref{eq-parti-jhamela}),
and \(H(\{r_i\}) = - \sum_i r_i \log_2 r_i\) is the 
Shannon entropy of a set of probabilities \(\{r_i\}\). Rewriting the capacity in terms of average occupation numbers, we obtain the 
classical capacity of 
channels carrying photons as
\begin{equation}
\label{semi-Qutab-Minar}
{\cal C}^{p}_{E} = - \sum_i \left[ \overline{n_i} \log_2 \overline{n_i} - 
\left( 1+ \overline{n_i} \right) \log_2 \left( 1+ \overline{n_i} \right) \right], 
\end{equation}
where \(\overline{n_i}\) is given by the Eq. (\ref{eq-ocuu-jhamela}).
Although we have obtained this capacity by maximizing  the Holevo quantity, 
the same result is also obtained by maximizing the  accessible information in the single-copy (nonasymptotic) case 
(see Eq. (\ref{kachuri1_hanuman_haluikar})).

The above classical capacity is obtained by using the average energy constraint. Several other types of constraints on the energy are possible 
and all of them give the same classical capacity to lowest order (see Ref. \cite{rmp}).

\section{Channels that carry massive indistinguishable particles} 
\label{sec-mota}

Let us now move on to the case of a communication channel that carries massive indistinguishable particles \cite{amader}. 
In this case, it is natural to impose the average particle number constraint along with the average energy constraint. 
The classical capacity in this case is the maximum of the Holevo quantity over all ensembles that satisfy both these 
constraints, while the quantum ones are equal to the maximum of the  entropy of the state that satifies both these constraints. 
As we will see, both classical and quantum capacities are equal for the noiseless channel.

A system of noninteracting massive bosons exhibits a Bose Einstein condensation: Below a certain critical temperature, 
the number of particles in the lowest energy level (ground state) becomes a significant fraction of what
is present in the whole system. 
In Ref. \cite{amader}, we considered the channel capacity in Eq. (\ref{khirish}) below, and showed that 
it can exhibit the onset of the condensation. Here we present more details of the calculations presented there. 
The figures presented in Ref. \cite{amader} were corresponding to the case where the bosons were confined to a three-dimensional cubic box
with periodic boundary conditions. Here we consider the harmonic trap, which is actually more close to usual experimental 
procedures. Unless mentioned otherwise, we consider in this paper a harmonic trap in three dimensions.

Let us note that noninteracting photons, despite being bosons, do not show condensation, due to particle number nonconservation. 
However, effectively interacting photons may acquire an effective mass, and show condensation (see for instance Ref. \cite{alor-jol}).

\subsection{Noninteracting bosons}
\label{sec-mutki}

Suppose that a quantum 
channel transmits spinless noninteracting bosons described by the Hamiltonian \({\cal H}_b = \sum_i \epsilon_i a^\dagger_i a_i\), and 
that they are confined to a trap of energy levels \(\{\epsilon_i\}\). 
We will begin by considering the classical capacity.

\subsubsection{Classical capacity}
\label{natun-natun-na-korda}

For a system whose average energy and average particle number
are given by \(E\) and \(N\), the allowed ensembles \(\{p_i, \rho_i\}\) must satisfy the average energy constraint  
\begin{equation}
\label{daktar-redo}
\mbox{tr}({\cal H}_b \overline{\rho}) = E,
\end{equation}
and the average particle number constraint
\begin{equation}
\label{daktar-knuti}
\mbox{tr}({\cal N} \overline{\rho}) = N,
\end{equation}
where \({\cal N} = \sum_i a^\dagger_i a_i\). 
The 
classical capacity given by 
\begin{equation}
{\cal C}_{E,N}^{be} = \max_{\{p_i, \rho_i\} \mbox{\footnotesize{ satisfying Eqs. (\ref{daktar-redo}) and (\ref{daktar-knuti}) }}}
\chi (\{p_i, \rho_i\}).
\end{equation}
Therefore,
\begin{equation}
\label{bostomi}
{\cal C}_{E,N}^{be} \leq \max_{\{p_i, \rho_i\} \mbox{\footnotesize{ satisfying Eqs. (\ref{daktar-redo}) and (\ref{daktar-knuti}) }}}
S(\overline{\rho}).
\end{equation}
Once again, the maximization of the rhs of inequality (\ref{bostomi}) was considered before, in Sec. \ref{sec-danguli}. 
There, it was discussed that the maximization of the rhs is attained by the grand canonical ensemble. 
The grand canonical ensemble consists of  number states, which are mutually orthogonal, and therefore, the 
channel capacity is attained for that ensemble, so that 
\begin{equation}
\label{khirish}
{\cal C}^{be}_{E,N} = H\left(\left\{ \frac{1}{{\cal Z}_{GC}^b}\exp\left(-\beta \sum_i(\epsilon_i - \mu^b)n^b_i  \right) \right\}\right).
\end{equation}
One may rewrite the capacity in terms of the average occupation numbers as (compare with Eq. (\ref{semi-Qutab-Minar}))
\begin{equation}
\label{Qutab-Minar}
{\cal C}^{be}_{E,N} = - \sum_i \left[ \overline{n_i^b} \log_2 \overline{n_i^b} - 
\left( 1+ \overline{n_i^b} \right) \log_2 \left( 1+ \overline{n_i^b} \right) \right].
\end{equation}
Again, the classical capacity obtained here by maximizing the Holevo quantity is the same as that obtained by maximizing 
the accessible information in the single-copy (nonasymptotic) case (see Eq. (\ref{kachuri1_hanuman_haluikar})).


\subsubsection{Quantum capacity}

We now consider the case of quantum capacity. Consider the noiseless quantum channel that 
transmits arbitrary pure states of \({\cal H}^{\otimes n}\), such that the energy and number of particles (and 
\emph{not} their average values) in any such 
pure state is \(nE\) and \(nN\) respectively. (Mixtures of such pure states are also noiselessly transmitted, by linearity.)
For a noiseless channel, the average fidelity condition is automatically satisfied (see Eq. (\ref{Shor_bhodrolok_chhilen_bhag-gish})). 
Consequently, the corresponding quantum capacity, \(Q_n\), irrespective of whether there is any classical side channel 
(see Sec. \ref{quantumcapa}), is given by 
\[
Q_n = \log_2 \Omega(nE,nN),
\]
where \(\Omega(nE,nN)\) is the dimension of the 
subspace spanned by pure states of \({\cal H}^{\otimes n}\) having energy \(nE\) and number of particles \(nN\).
So, we are now in the microcanonical ensemble. (To be more precise, one should allow for small fluctuations around \(nE\) and \(nN\), but the derivation is similar.)

The system (described on the Hilbert space \({\cal H}^{\otimes n}\)) consists of \(n\) identical parts, each being 
described on the Hilbert space \({\cal H}\). Since the parts are identical, each part, \emph{on average}, has energy \(E\) and number of  
particles \(N\). Consider one such part, and let \(l(\epsilon,m)\) be the dimension of the subspace of that part whose 
energy is \(\epsilon\), and which consists of \(m\) particles. (Again, we are disregarding fluctuations.)
Then,  
\begin{widetext}
\begin{eqnarray}
\log_e \Omega(nE,nN) &=& \log_e \sum_{\epsilon, m} l(\epsilon, m) \Omega(nE-\epsilon, nN-m) \nonumber \\
&=& \log_e \sum_{\epsilon, m} l(\epsilon, m) \exp\left[\log_e\Omega((n-1)E +E-\epsilon, (n-1)N +N-m)\right], \nonumber
\end{eqnarray}
\end{widetext}
where we have 
\[
\left| \frac{E-\epsilon}{(n-1)E}\right| \ll 1, \quad \left| \frac{N-m}{(n-1)N}\right| \ll 1,
\]
as we suppose that \(n\) is large (compare with e.g. \cite{Sona-Jana-ebong-Khurdo}).
Therefore we can expand \(\log_e\Omega((n-1)E +E-\epsilon, (n-1)N +N-m)\)
as 
\begin{widetext}
\begin{eqnarray}
\label{tumi-je-a(n)adhar-tai-boro-bhalobasi}
\log_e \Omega((n-1)E +E-\epsilon, (n-1)N +N-m) = \log_e \Omega((n-1)E, (n-1)N) \nonumber \\
+ \frac{\partial \log_e \Omega}{\partial E}\Big|_{E=(n-1)E, N=(n-1)N} (E- \epsilon)
+ \frac{\partial \log_e \Omega}{\partial N}\Big|_{E=(n-1)E, N=(n-1)N} (N-m) + \ldots .
\end{eqnarray}
Therefore,
\begin{eqnarray}
\log_e \Omega(nE,nN) &=& \log_e \Omega ((n-1)E, (n-1)N) + \log_e \sum_{\epsilon,m} l(\epsilon,m) \mbox{e}^{-\beta \epsilon + \mu^b \beta m} + \beta E - \mu^b \beta N \nonumber \\
&=& \log_e \Omega ((n-1)E, (n-1)N) + \log_e {\cal Z}^b_{GC}(E,N) + \beta E - \mu^b \beta N, \nonumber
\end{eqnarray}
\end{widetext}
where \({\cal Z}^b_{GC}(E,N)\) is the grand canonical partition function of the system  described on \({\cal H}\), and having 
average energy \(E\) and average particle number \(N\).
We will ultimately be interested in dividing the logarithm of \(\Omega(nE,nN)\) by \(n\) and consider the limit as 
\(n\to\infty\), and in that case, the further terms on the right-hand-side 
of Eq. (\ref{tumi-je-a(n)adhar-tai-boro-bhalobasi}), after the third term, 
will not contribute. 
We now use the ``fundamental equation for open systems'', 
\[
-k_B T \log_e {\cal Z}^b_{GC} = E -k_B T (\log_e 2) S(E,N) - \mu^b N,
\]
where \(S(E,N)\) is the von Neumann entropy of the grand canonical state of the system 
described on \({\cal H}\), and having 
average energy \(E\) and average particle number \(N\). 
(Note that we are using the information theoretic definition of entropy as in Eq. (\ref{amar-byala-je-jai}), in which 
the multiplicative constant \(k_B\) is absent, and the logarithm is to the base 2.) 
Finally therefore,
\[
\log_2 \Omega(nE,nN) = \log_2 \Omega ((n-1)E, (n-1)N) + S(E,N).
\]
Similarly,
\begin{eqnarray}
&&\log_2 \Omega((n-1)E, (n-1)N) \nonumber \\ 
&&= \log_2 \Omega ((n-2)E, (n-2)N) + S(E,N), \nonumber 
\end{eqnarray}
so that 
\[
\log_2 \Omega(nE,nN) = \log_2 \Omega ((n-2)E, (n-2)N) + 2S(E,N).
\]
This recursion can be carried on for \(p\) times, if \(n-p\) is large \cite{bayonet-hok-joto-dharalo-kaste-ta-shan-diyo-bondhu}, 
so that we have 
\begin{equation}
\label{robimama-dei-hama-gaye-ranga-jama-oi}
\log_2 \Omega(nE,nN) = \log_2 \Omega ((n-p)E, (n-p)N) + pS(E,N).
\end{equation}
We moreover demand that \(\lim\limits_{n \to \infty} \frac{p}{n} = 1\), which can be satisfied even if \(n-p\) is large 
\cite{Shell-aar-bomb-hok-joto-bharalo-kaste-ta-shan-diyo-bondhu}.
Since \(n-p \ll n\) \cite{bou-aar-bor-e-ekhon-jhogrda-cholchhey}, we have 
\[\lim_{n \to \infty} \frac{1}{n} \log_2 \Omega ((n-p)E, (n-p)N) =0,\] 
so that 
\[
\lim_{n \to \infty}\frac{\log_2 \Omega(nE,nN)}{n} = S(E,N).
\]


Therefore, the quantum capacity, \(Q^{be}_{E,N}\), irrespective of whether 
we have an additional side channel for classical information transfer, 
of a noiseless quantum channel that transmits arbitrary states, \(\rho\), of \({\cal H}\), such that 
the average energy \(\mbox{tr}({\cal H}_b \rho) = E\) and the average number of particles 
\(\mbox{tr}({\cal N} \rho) = N\), is given by 
\[
Q^{be}_{E,N}= \lim_{n \to \infty}\frac{\log_2 \Omega(nE,nN)}{n} = S(E,N).
\]
We therefore see that the quantum capacity is  equal to the classical capacity obtained in Sec. \ref{natun-natun-na-korda}.


\subsubsection{Spinless bosons in a harmonic trap}

\begin{figure}[tbp]
\begin{center}
\epsfig{figure= 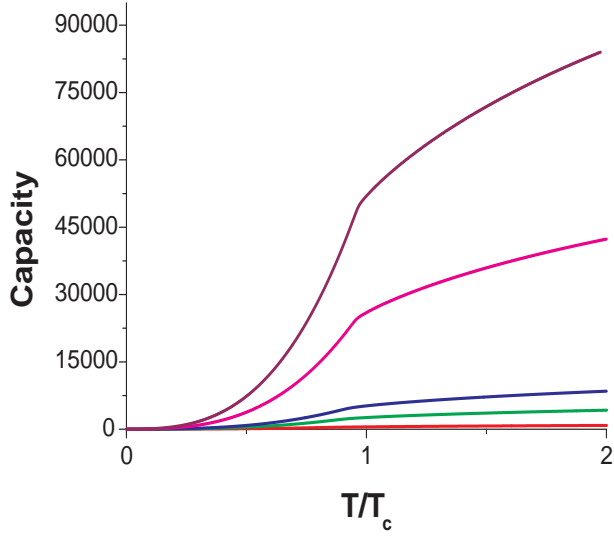,height=.35\textheight,width=0.5\textwidth}
\caption{(Color online)  Capacity (in bits or qubits) as function of the dimensionless variable 
\(T/T_c\) for different number of bosons. At the critical temperature, 
the capacity changes its behavior from
being concave to being convex (with respect to temperature). 
Again, the curves (from bottom to top) 
 represent  100, 500, 1000, 5000, and 10000 bosons, respectively. 
}
\label{fig_cap_diffN}
 \end{center}
\end{figure}

Let us now consider the case of noninteracting massive spinless bosons confined in a harmonic trap. The energy levels are therefore
\begin{equation}
\hbar  (\omega_x n_x + \omega_y n_y + \omega_z n_z), \quad n_x, n_y, n_z = 0, 1, 2, \ldots.
\end{equation}
In evaluating the capacity numerically, we must cut off the infinite sequence of energy levels at some point. The evaluated
``capacity'' of course depends on the point of the cut-off.
\begin{figure}[tbp]
\begin{center}
\epsfig{figure= 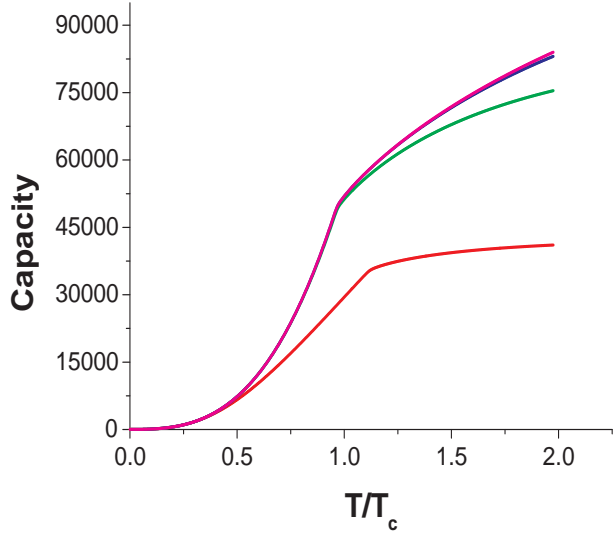,height=.35\textheight,width=0.5\textwidth}
\caption{(Color online) Capacity (in bits or qubits) as a function of the dimensionless variable
 \(T/T_c\) for  \(N = 10^4\) massive bosons in a harmonic trap,
with different summation boundaries. $T_c$ denotes the critical temperature in the thermodynamic 
limit. Therefore, \(T_c = \frac{\hbar \omega_{ho}}{k_B} \left(\frac{N}{\zeta(3)}\right)^{1/3}\), 
where \(\omega_{ho} = \left(\omega_x \omega_y \omega_z \right)^{1/3}\) is the geometrical average of the frequencies and 
\(\zeta\) is the Riemann zeta function.  For the purpose of all the figures in this paper, we consider a harmonic trap, and 
 the case when 
\(\omega_x = \omega_y = \omega_z\). The four curves (from bottom to top) 
represent cut-offs at \(n_x = n_y = n_z =\) 40, 120, 220, and 300, respectively. 
Note that the last two curves  almost coincide. 
}
\label{fig_SvsTcutoff}
 \end{center}
\end{figure}
In Fig. \ref{fig_SvsTcutoff}, we plot the channel capacity for the case of \(10^4\) bosons (with respect to temperature), 
where we show the ``capacities'' for the different  
points of the cut-off. In this case, the convergence is obtained roughly at about \(n_x=n_y=n_z =300\). We see that the curves show a 
sort of fracture at a certain temperature, changing their behavior from being 
concave to being convex with respect to temperature. 
\begin{figure}[tbp]
\begin{center}
\epsfig{figure= 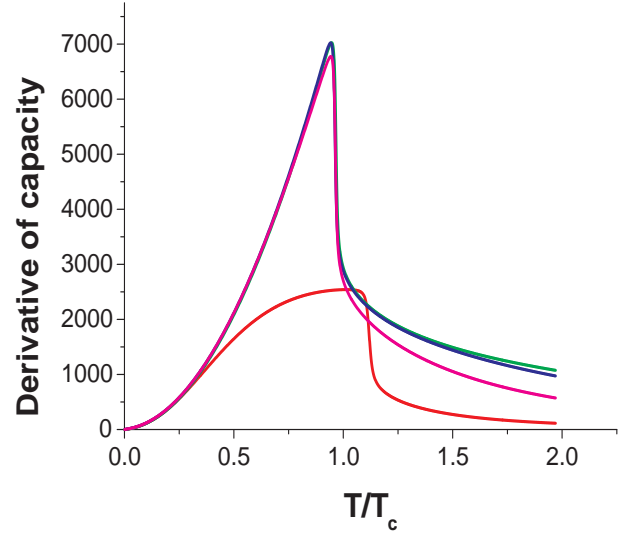,height=.35\textheight,width=0.5\textwidth}
\caption{(Color online) Derivative of the capacity (in bits or qubits) as function of the dimensionless variable 
\(T/T_c\) for \(N = 10^4\) bosons. Note that the number of bosons is the same as that in the 
previous figure.
Moreover, the  curves in this figure (from bottom to top) 
 represent the cut-offs at 40, 120, 220, and 300,  respectively, which are 
exactly the same as those in the previous figure. 
Note here also that the last two curves  almost coincide. 
}
\label{fig_derivative_cutoff}
 \end{center}
\end{figure}
This can be seen more clearly in Fig. \ref{fig_derivative_cutoff}, where we plot the derivative of the 
plots in the preceding 
figure. For calculating the derivative, we use a four-point formula:
\begin{eqnarray}
\frac{df}{dx} &=& \frac{1}{12 \Delta x} \Big[  f(x - 2\Delta x) - 8f(x - \Delta x) + 8 f(x + \Delta x) \nonumber \\
 && - f(x + 2 \Delta x) \Big]
+ O(\Delta x^4).
\end{eqnarray}
In the thermodynamic limit, this bending causes the derivative to have a discontinuity, and the corresponding 
temperature is the critical temperature of the condensation. 
\begin{figure}[tbp]
\begin{center}
\epsfig{figure= 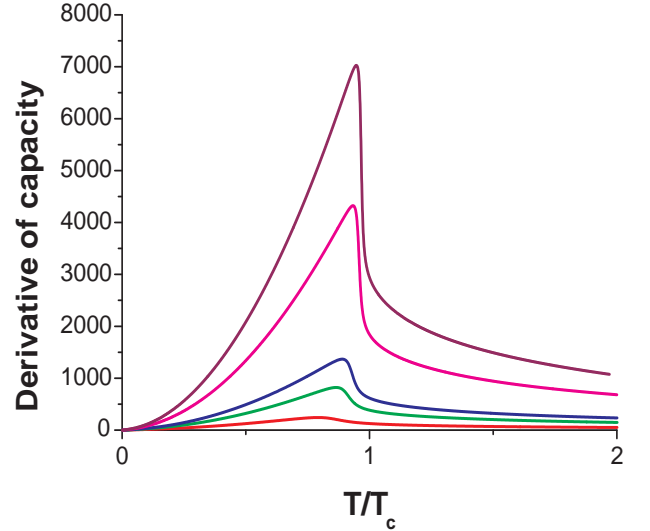,height=.35\textheight,width=0.5\textwidth}
\caption{(Color online)  Derivative of the capacity (in bits or qubits) as function of the dimensionless variable 
\(T/T_c\) for different number of bosons.
Curves (from bottom to top) 
represent  100, 500, 1000, 5000, and 10000 bosons, respectively.
}
\label{fig_deri_cap_diffN}
 \end{center}
\end{figure}
Fig. \ref{fig_deri_cap_diffN} shows the development of this discontinuity in the derivative of the capacity,
with increasing average total particle number \(N\). The corresponding capacities are plotted in Fig. \ref{fig_cap_diffN}.
As seen in Figs.  \ref{fig_deri_cap_diffN} and  \ref{fig_cap_diffN}, the critical 
temperature corresponding to the fracture grows with \(N\), as expected. 
The fractures indicate the onset of the condensation for the corresponding values of \(N\). 
The gap between the temperature corresponding to the fracture and the thermodynamic critical temperature 
is known to exist, and estimated values of the gap has also been given (see Ref. \cite{Dalfovo}). 
Also, the capacities grow with the number of particles, as expected.

\begin{figure}[tbp]
\begin{center}
\epsfig{figure= 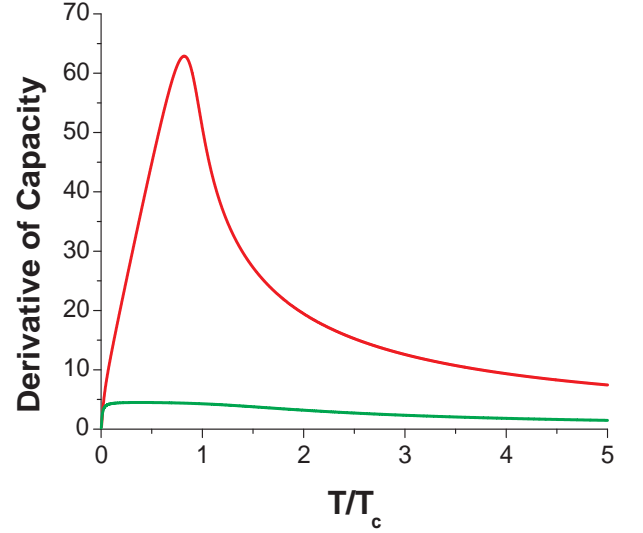,height=.35\textheight,width=0.5\textwidth}
\caption{(Color online)  Derivatives of the capacities (in bits or qubits) are plotted as functions of the
 temperature for \(N = 100\) bosons in one- and two-dimensional traps. 
The upper curve represents the 2D case, where the horizontal axis is of the dimensionless variable \(T/T_c^{2D}\).
And \(T_c^{2D} = \frac{\hbar \omega_{ho}^{2D}}{k_B} \left(\frac{N}{\zeta(2)}\right)^{1/2}\), where
\(\omega_{ho}^{2D} = \left(\omega_x \omega_y\right)^{1/2}\) \cite{Dalfovo}. For the figure, we consider the case when
\(\omega_x = \omega_y\).
In the one-dimensional case (lower curve), the horizontal axis is of the dimensionless variable \(T/T_c^{1D}\), where
\(T_c^{1D} = \frac{\hbar \omega_{ho}}{k_B} \frac{N}{\log_e (2N)}\) \cite{Ketterle96}. It is clear 
from the figure that capacity of a one-dimensional trap does not show any onset of condensation, while that for a 
two-dimensional trap clearly does show it.}
\label{fig_1Dvs2D}
 \end{center}
\end{figure}

It is known that the dimension of the system under consideration plays a role in determining whether a condensation exists.
For example, in the case of harmonic traps, the 3D and 2D traps exhibits condensations, while the 1D case does not show a condensation 
(see e.g. \cite{Dalfovo}). The cases considered in the previous (and latter) figures 
are 3D harmonic traps. In Fig. \ref{fig_1Dvs2D}, we compare the 
qualitative behavior of the derivatives of 
the capacities for 2D and 1D. 

Going back to the case of the 3D harmonic trap, we consider energy as a function of the temperature, and see that the criticality is 
clearly visible in these curves. 
\begin{figure}[tbp]
\begin{center}
\epsfig{figure= 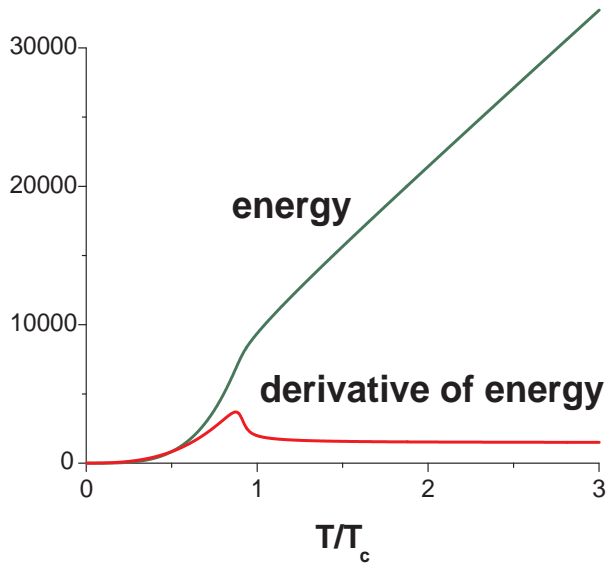,height=.35\textheight,width=0.5\textwidth}
\caption{(Color online)  For \(N =500\) bosons, the dimensionless variable
\(\mbox{energy}/\hbar \omega\) (where \(\omega_x = \omega_y =\omega_z = \omega\)) 
and its derivative,
are plotted against the dimensionless variable \(T/T_c\). 
Energy behaves quite similarly as the capacity, in that  
it changes its behavior from being convex to being concave, with respect to temperature, at the  criticality. 
This is more clear from the  derivative of the energy. 
}
\label{fig_en_derien_500}
 \end{center}
\end{figure}
In Fig. \ref{fig_en_derien_500}, we plot the energy and
its derivative with respect to temperature,  for 500 spinless bosons in a harmonic trap. Note that 
the capacity plotted with respect to energy does not indicate the condensation.

\subsection{Noninteracting fermions in a harmonic trap}

Let us now move over to the case of noninteracting fermions. 
Similar arguments as in the preceding subsection imply that the capacities of \(N\) spin-\(s\) noninteracting fermions in a 
trap with energy levels \(\{\epsilon_i\}\) are given by
\begin{eqnarray}
\label{shirish}
&& {\cal C}^{fd}_{E,N} = Q^{fd}_{E,N} = \nonumber \\
&& H\left(\left\{ \frac{1}{{\cal Z}_{GC}^f}\exp\left(-\beta \sum_i \sum_{\mbox{\footnotesize{mag}}=-s}^{s}
(\epsilon_i - \mu^f)n^f_{i,\mbox{\footnotesize{mag}}}  \right) \right\}\right),\nonumber \\
\end{eqnarray}
where 
\begin{equation}
{\cal Z}_{GC}^f = \prod_i \left[ 1 +  \exp\left(-\beta (\epsilon_i - \mu^f)  \right)            \right]^g,
\end{equation}
the power \(g = 2s+1\) being present due to the degeneracy of the spin states.  
The suffix ``mag'' represents the magnetic quantum number.
\(\mu^f\) represents the chemical potential in this case.
The channel capacity in this case is reached by the fermionic grand canonical ensemble
\begin{eqnarray}
\Big\{\frac{1}{{\cal Z}^f_{GC}} \exp\left(-\beta \sum_i 
\sum_{\mbox{\footnotesize{mag}}=-s}^{s}
(\epsilon_i - \mu^f) n_{i,\mbox{\footnotesize{mag}}}^f\right), \nonumber \\
\left|n_{0,-s}^f, n_{0,-s+1}^f, \ldots, n_{0,s}^f, n_{1,-s}^f, n_{1,-s+1}^f, \ldots, n_{1,s}^f,  \ldots
\right\rangle\Big\},\nonumber \\
\end{eqnarray}
 where the elements of the ensemble runs over all 
combinations of the \(n_{i,\mbox{\footnotesize{mag}}}^f\)'s (\(n_{i,\mbox{\footnotesize{mag}}}^f = 0,1\), and mag \(= -s, -s +1, \ldots, s\)).
Again the capacities can be rewritten in terms of the average occupation numbers 
\begin{equation}
\overline{n_i^f} = \frac{g}{e^{\beta (\epsilon_i - \mu^f)} + 1},
\end{equation}
as
\begin{eqnarray}
&& {\cal C}^{fd}_{E,N} = Q^{fd}_{E,N}=  \nonumber \\
&& - g \sum_i \left[ \frac{\overline{n_i^b}}{g} \log_2 \frac{\overline{n_i^b}}{g} - 
\left( 1+ \frac{\overline{n_i^b}}{g} \right) \log_2 \left( 1+ \frac{\overline{n_i^b}}{g} \right) \right]. \nonumber \\
\end{eqnarray}

\begin{figure}[tbp]
\begin{center}
\epsfig{figure= 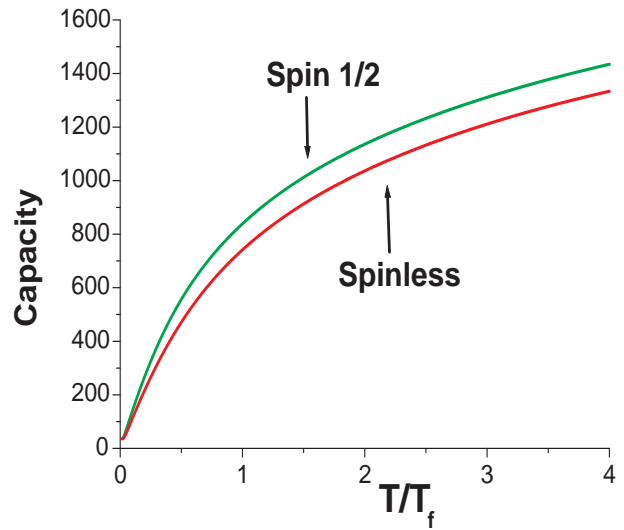,height=.35\textheight,width=0.5\textwidth}
\caption{(Color online)  For \(N =100\) fermions, the 
capacity (in bits or qubits) of spin-1/2 fermions is compared with that of spinless fermions. The horizontal axis is of the dimensionless variable \(T/T_f\), where 
\(T_f\) is the Fermi temperature for spin-1/2 fermions. Note that we use the same 
Fermi temperature for \emph{both} the curves. The Fermi temperature for a harmonic trap is given by 
\(T_f = \frac{\hbar \omega_{ho} }{k_B} \left(\frac{6N}{g}\right)^{1/3}\). 
}
\label{fig_spinless_vs_spinhalf}
 \end{center}
\end{figure}

In Fig. \ref{fig_spinless_vs_spinhalf}, we compare
the capacities of spinless fermions with that of spin 1/2 fermions in a harmonic trap.

Noninteracting fermions do not exhibit a condensation. However, interacting fermions can exhibit Cooper pairing, and consequently 
superfluid BCS  transition. The channel capacity in such case also indicates the onset of the BCS transition (see Ref. \cite{amader}).

\section{Fermions are better carriers of information than bosons}
\label{sec-ghyama}

Numerical simulation with several values of the total number of particles, with several types of traps, and over a large 
range of temperature reveals that fermions are better carriers of information than bosons. For sufficiently high temperatures, we 
have been able to obtain analytical results in this respect, as we have already presented in the following theorem from Ref. \cite{amader}. 
Here, we give a detailed proof of the theorem.
We have already proven that the classical and quantum capacities are the same in the case under study. For definiteness, we will
consider only the classical capacity in this section and the succeeding one.

\noindent \textbf{Theorem.} \emph{For power law potential traps (with power \(\gamma\) and dimension \(d\)), and for sufficiently high 
temperatures, the capacity of spinless fermionic channel is better than that of spinless bosonic channel when} 
\begin{equation}
\label{asli-cheez}
\frac{1}{\gamma} + \frac{1}{2} > \frac{1}{d}.
\end{equation}

Note that a power law potential with power \(\gamma\) and dimension \(d\) is given by 
\(V\left(x_1, x_2, \ldots, x_d\right) = r^\gamma\), where \(r = \sqrt{x_1^2 + x_2^2 + \ldots + x_d^2 }\), 
in the \(d\)-dimensional Cartesian 
space \(\left(x_1, x_2, \ldots, x_d\right)\).

\noindent \textbf{Proof.}
Let us start with the case of (spinless) bosons and perform the high temperature expansion. First we expand the fugacity 
\begin{equation}
z_b = \exp(\beta \mu^b), 
\end{equation}
in powers of \(N/S_1\), where 
\begin{equation}
\label{eq-S}
S_k = \sum_i \exp(-k\beta \epsilon_i), \quad k=1,2, \ldots.
\end{equation}
Note that 
\begin{eqnarray}
\label{Chandragupta}
N &=& \sum_i \overline{n_i^b} \nonumber \\
  &=& \sum_i \frac{1}{\frac{1}{z_b}e^{\beta \epsilon_i} - 1} \nonumber \\
  &=& \sum_i z_b \exp \left(- \beta \epsilon_i \right) \left[ 1 - z_b \exp \left(-\beta \epsilon_i \right) \right]^{-1}
		\nonumber \\
&=& z_b S_1 + z_b^2 S_2 + z_b^3 S_3 + \ldots.
\end{eqnarray}
Suppose now that 
\begin{equation}
\label{Maurya}
z_b = b_0 + b_1 \frac{N}{S_1} + b_2 \left(\frac{N}{S_1}\right)^2 
+ b_3 \left(\frac{N}{S_1}\right)^3 + \ldots.
\end{equation}
Substituting \(z_b\), as in Eq. (\ref{Maurya}), into Eq. (\ref{Chandragupta}), and comparing different powers of 
\(N/S_1\), we find that 
\begin{eqnarray}
b_0 = 0, & \quad & b_1 = 1, \nonumber \\
b_2 = - \frac{S_2}{S_1}, & \quad &
b_3 = \frac{2S_2^2 - S_1 S_3}{S_1^2}.
\end{eqnarray}
We are now in a position to expand \({\cal C}_{E,N}^{be}/\log_2 e\) in powers of \(N/S_1\). We use the 
expression in Eq. (\ref{Qutab-Minar}) for this purpose. In Eq. (\ref{Qutab-Minar}), the substitutions 
in the different terms are done as follows.
For the term  \(\overline{n_i^b} \log_e \overline{n_i^b}\), for  \(\overline{n_i^b}\),
 we substitute
\begin{eqnarray}
\overline{n_i^b} &=& z_b \exp \left(- \beta \epsilon_i \right) \left[ 1 - z_b \exp \left(-\beta \epsilon_i \right) \right]^{-1}
	\label{Humayun}\\
&=& z_b e^{- \beta \epsilon_i } + z_b^2 e^{- 2\beta \epsilon_i } + z_b^3 e^{- 3\beta \epsilon_i}
	+ \ldots. \label{Jahangir}
\end{eqnarray}
and for \(\log_2 \overline{n_i^b}\), we use Eq. (\ref{Humayun}) to write
\begin{eqnarray}
\log_e \overline{n_i^b} &=& 
\log_e z_b - \beta \epsilon_i - \log_e \left( 1 - z_b \exp \left(-\beta \epsilon_i \right) \right) \nonumber \\
&=& \log_e z_b - \beta \epsilon_i \nonumber \\
&& + \left( z_b e^{- \beta \epsilon_i} + \frac{1}{2} z_b^2 e^{- 2\beta \epsilon_i} +
\frac{1}{3} z_b^3 e^{- 3\beta \epsilon_i} + \ldots \right). \nonumber \\
\end{eqnarray}
and this is then used for the substitution.
For the term \(\left(1 + \overline{n_i^b}\right) \log_e \left( 1+ \overline{n_i^b} \right)\), 
we use Eq. (\ref{Jahangir}) to substitute for \(\overline{n_i^b}\) in both places, 
and then \(\log_e\left(1+\overline{n_i^b}\right)\)
is expanded in powers of \(\overline{n_i^b}\).

A similar calculation is done for spinless fermions. In this case, the fermionic fugacity 
\begin{equation}
z_f = \exp\left( \beta \mu^f \right)
\end{equation}
can be expanded as 
\begin{equation}
z_f = f_0 + f_1 \frac{N}{S_1} + f_2 \left(\frac{N}{S_1}\right)^2 
+ f_3 \left(\frac{N}{S_1}\right)^3 + \ldots,
\end{equation}
in which the \(f_i\)'s are obtained from the average particle number conservation equation 
\begin{eqnarray}
N &=& \sum_i \overline{n_i^f} \nonumber \\
  &=& \sum_i z_f \exp \left(- \beta \epsilon_i \right) \left[ 1 + z_f \exp \left(-\beta \epsilon_i \right) \right]^{-1}
	\nonumber \\
&=& z_f S_1 - z_f^2 S_2 + z_f^3 S_3 - \ldots
\end{eqnarray}
as 
\begin{eqnarray}
f_0 = 0, & \quad & f_1 = 1, \nonumber \\
f_2 =  \frac{S_2}{S_1}, & \quad &
f_3 = \frac{2S_2^2 - S_1 S_3}{S_1^2}.
\end{eqnarray}
Note that these are the same as in the case of bosons, except for the sign in \(f_2\).

We perform the calculation up to the third order, and find that 
\begin{equation}
\label{eq-bose-capa}
\frac{{\cal C}^{be}_{E,N}}{\log_2 e} = \sum_{i=1}^{3} \alpha^b_i \left(\frac{N}{S_1}\right)^i + 
\beta^b_1 \frac{N}{S_1} \log_e \frac{N}{S_1} + \beta^b_2 \left(\frac{N}{S_1}\right) ^2\log_e \frac{N}{S_1}
\end{equation}
plus  higher order terms,
whereas 
\begin{equation}
\label{eq-fermi-capa}
\frac{{\cal C}^{fd}_{E,N}}{\log_2 e} = \sum_{i=1}^{3} \alpha^f_i \left(\frac{N}{S_1}\right)^i + 
\beta^f_1 \frac{N}{S_1} \log_e \frac{N}{S_1} + \beta^f_2 \left(\frac{N}{S_1}\right) ^2\log_e \frac{N}{S_1}
\end{equation} 
plus higher order terms.
The coefficients of first order perturbation are equal: 
\begin{equation}
\alpha_1^b = \alpha_1^f = S_1 + D_1,
\end{equation} 
where we have set 
\begin{equation}
\label{eq-D}
D_k = \sum_i \beta \varepsilon_i \exp\left(- k \beta \varepsilon_i\right), \quad k=1,2, \ldots.
\end{equation}
In the next order, they differ by a sign: 
\begin{equation}
\label{eq-alpha_DS}
\alpha_2^b = - \alpha_2^f = \frac{S_2}{2} - \frac{S_2}{S_1} D_1 + D_2.
\end{equation} 
The third order perturbation coefficients are again equal: 
\begin{equation}
\alpha_3^b = \alpha_3^f = - 3S_2 + \frac{S_3}{3} + \frac{2S_2^2}{S_1} + \frac{2S_2^2 - S_1 S_3}{S_1^2} D_1
- \frac{2S_2}{S_1} D_2 + D_3. 
\end{equation}
Also, 
\begin{eqnarray}
\beta_1^b = \beta_1^f = - S_1, \nonumber \\
\beta_2^b =0, \phantom{a} \beta_2^f = 2S_2.
\end{eqnarray}


Now, upto third order, the only coefficients that are different are those of \((N/S_1)^2\) and
\((N/S_1)^2 \log_e (N/S_1)\).
We have 
\begin{equation}
0 = \beta_2^b < \beta_2^f.
\end{equation}
Also 
\begin{equation}
\label{eq-impo}
\alpha_2^b = S_2 \left( \frac{1}{2} - \beta \frac{D_1}{S_1}  + \beta  \frac{D_2}{S_2} \right).
\end{equation}
We now evaluate \(\frac{D_1}{S_1}\) and \(\frac{D_2}{S_2}\) by using  the density of states \cite{appu}
\begin{equation}
\label{Srikant}
\varsigma(\epsilon) = \int d^d p d^d x \phantom{a} \delta \left(\frac{p^2}{2m} + V(\vec{r}) - \epsilon \right)
\end{equation}
for a system in \(d\) dimensionsal Cartesian space \(\vec{r} = (x_1, x_2, \ldots, x_d)\), trapped in a potential
\(V(\vec{r})\). \(\vec{p} = (p_1, p_2, \ldots, p_d)\) denotes the respective momentum, with \(p = |\vec{p}|\).  
The integration in Eq. (\ref{Srikant}) is over the phase space. We may rewrite \(\varsigma(\epsilon)\) as 
\cite{aaj-bikeley-ki-bhai-h(n)at_tey-jabi-na-jabina}
\begin{equation}
\label{Niren-Chattopadhyay}
\varsigma(\epsilon) = \mbox{constant} \cdot \int dx_1 \ldots dx_d (\epsilon -V)^{(d-2)/2}.
\end{equation}

Let us now consider the potential as \cite{appu}
\begin{equation}
\label{eq-power}
V(\vec{r}) = r^\gamma,
\end{equation}
where \(r = \sqrt{x_1^2 + \ldots + x^2_d}\). In this case, 
\begin{eqnarray}
\varsigma(\epsilon) &=& \mbox{constant} \cdot \quad \epsilon^{\frac{d}{\gamma} + \frac{d-2}{2}} \nonumber \\
                       &&\times \int d\tilde{x}_1 \ldots d\tilde{x}_d 
\left( 1 - \left( \tilde{x}_1^2 + \ldots+ \tilde{x}_d^2 \right)^{\gamma/2} \right)^{(d-2)/2},\nonumber
\end{eqnarray} 
so that 
\begin{eqnarray}
\varsigma(\epsilon) = \mbox{constant} \cdot \quad \epsilon^{\frac{d}{\gamma} + \frac{d-2}{2}},
\end{eqnarray} 
where  \(x_i =\epsilon^{\frac{1}{\gamma}}  \tilde{x}_i\) (\(i= 1, \ldots, d\)).  We are now ready to
calculate the sums in  Eq. (\ref{eq-impo}).  We have 
\begin{eqnarray}
\frac{D_1}{S_1} & = &\frac{\int \beta \epsilon \exp(- \beta \epsilon) \varsigma(\epsilon) d \epsilon}
{\int  \exp(- \beta \epsilon) \varsigma(\epsilon) d \epsilon} \nonumber \\
 &= & \frac{\int \beta \epsilon \exp(- \beta \epsilon) \epsilon^{\frac{d}{\gamma} + \frac{d-2}{2}} d \epsilon}
{\int \exp(- \beta \epsilon) \epsilon^{\frac{d}{\gamma} + \frac{d-2}{2}} d \epsilon}  \nonumber  \\
& = & \frac{\Gamma(\frac{d}{\gamma} + \frac{d-2}{2} +1)} {\Gamma(\frac{d}{\gamma} + \frac{d-2}{2})}  \nonumber \\
&= & \frac{d}{\gamma} + \frac{d}{2}.
\end{eqnarray}
The integrations over \(\epsilon\) are performed from \(0\) to \(\infty\). Similarly, one can calculate \(\frac{D_2}{S_2}\), and it is equal to 
\(\frac{1}{2}(\frac{d}{\gamma} + \frac{d}{2})\). Therefore,
\begin{equation}
\alpha_2^b = \frac{S_2}{2} \left[ 1- \left(\frac{d}{\gamma} + \frac{d}{2} \right) \right].
\end{equation}
Therefore,
\begin{equation}
\alpha_2^b \leq 0 \leq \alpha_2^f
\end{equation}
holds, when 
\begin{equation}
\frac{1}{\gamma} + \frac{1}{2} > \frac{1}{d}.
\end{equation}
That is, the fermionic capacity is greater than the bosonic one in such cases. However, 
we must now check for which potentials and dimensions, the above perturbation technique is systematic.

In the expansion of the capacity in terms of \(N/S_1\), given in Eq. (\ref{eq-bose-capa}) for bosons, and in 
Eq. (\ref{eq-fermi-capa}) for fermions,  the coefficients of \(\frac{N}{S_1}\), \(\frac{N}{S_1} \log_e \frac{N}{S_1}\), 
\((\frac{N}{S_1})^2\), \((\frac{N}{S_1})^2 \log_e \frac{N}{S_1}\), and \((\frac{N}{S_1})^3\), 
are all of the order \(\beta ^{-(\frac{d}{\gamma} + \frac{d}{2})}\), since 
\begin{eqnarray}
S_1 & =&  \int \exp(- \beta \epsilon) \epsilon^{-(\frac{d}{\gamma} + \frac{d-2}{2})} d\epsilon \nonumber \\
      &\sim &  \beta ^{-(\frac{d}{\gamma} + \frac{d}{2})},
\end{eqnarray}
and 
\begin{eqnarray}
D_1 & =&  \int \beta \epsilon \exp(- \beta \epsilon) \epsilon^{-(\frac{d}{\gamma} + \frac{d-2}{2})} d\epsilon \nonumber \\
     & \sim & \beta ^{-(\frac{d}{\gamma} + \frac{d}{2})}.
\end{eqnarray}
Moreover, \(\frac{N}{S_1} \sim \beta^{(\frac{d}{\gamma} + \frac{d}{2})}\).
For systematics of the expansion in Eq. (\ref{eq-bose-capa}) for bosons, we 
need that the orders of \(\beta\) in  \(\alpha_1^b \frac{N}{S_1}\), \(\beta_1^b \frac{N}{S_1} \log_e \frac{N}{S_1}\), 
\(\alpha_2^b (\frac{N}{S_1})^2\), and \(\alpha_3^b (\frac{N}{S_1})^3\) should be in  increasing order of \(\beta\). This demand
leads to the following condition:
\begin{equation}
\label{eq-system}
\frac{2d}{\gamma} + d > \frac{d}{\gamma} + \frac{d}{2} > 1 > 0.
\end{equation}
This requires that 
\(\frac{1}{\gamma} + \frac{1}{2} > \frac{1}{d}\), which is the same as the condition required for
 fermions having a higher capacity than bosons. 
Similar calculation for the fermions leads to the same requirement. 


Lastly, note that \(N/S_1 \) tends to zero implies that \(T \rightarrow \infty\). 
This completes the proof. \(\square\)

\noindent \textbf{Remark 1.} The condition in Eq. (\ref{asli-cheez}) includes e.g. the harmonic trap in 2D and 3D, the 3D rectangular box, 
and the 3D 
spherical box.

\noindent \textbf{Remark 2.} In the proof, we work up to order \((N/S_1)^3\), and so the theorem holds for quite moderate temperatures. 

\begin{figure}[tbp]
\begin{center}
\epsfig{figure= 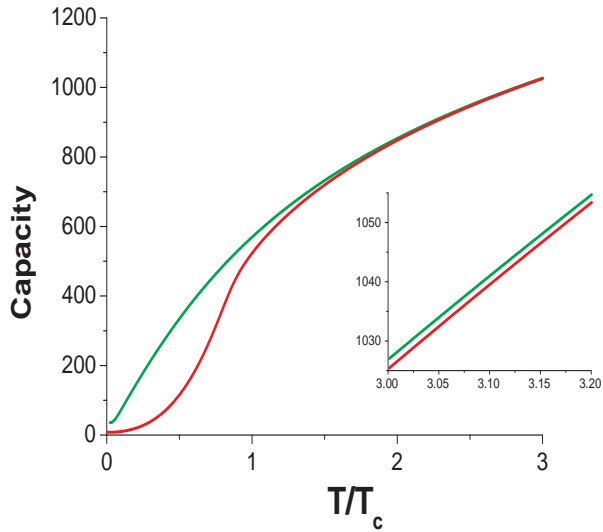,height=.35\textheight,width=0.5\textwidth}
\caption{(Color online)  Bosons vs. Fermions. We plot the capacities (in bits or qubits) 
of spinless bosons and spinless fermions with respect to the dimensionless variable \(T/T_c\), for 100 particles. 
Note that the horizontal axis represents \(T/T_c\) for \emph{both} the curves. The lower curve is for bosons. 
The figure clearly agrees with our theorem. Moreover, this figure along with other numerical simulations that we have performed,
indicates that the theorem is true even for low temperatures. The inset compares the behavior in an exemplary case at high temperatures. 
}
\label{fig_bosonsvsfermions}
 \end{center}
\end{figure}

At this point, please note that  we have numerically checked that the statement of the 
theorem holds also for low temperatures for the harmonic trap 
and for the 3D rectangular box \cite{amader}. In Fig. 
\ref{fig_bosonsvsfermions}, we compare the capacities of channels carrying 
bosons and fermions for \(N= 100\) for the case of a harmonic trap. 





\section{Low temperature behavior}
\label{sec-jotota-bhebechhilum-totota-soja-noi}

As we have already stressed, although the low temperature behavior of the capacity is not covered by the \textbf{Theorem} in the preceding section, 
numerical simulations seems to indicate that the statement of the \textbf{Theorem} is indeed true for lower temperatures. 
In this section, we want to point out that the statement can be proven analytically in some circumstances, for low temperatures.
For example, for the case of a bounded volume \(V\), containing \(N\) 
particles at temperature \(T\),
 if the particles are bosons, then the capacity at low temperatures (specifically, for 
\(T \leq T_c\), where in this case, \(T_c = \hbar^2 \frac{2\pi}{mk_B} \left( \frac{N}{V\zeta(3/2)} \right)^{\frac{2}{3}}\)) is given by 
(see e.g. \cite{molla-nasiruddin})
\[
\frac{C^{be}_{V,N,T}}{\log_2 e} = \frac{5V\zeta(5/2)}{2\hbar^3} \left( \frac{mk_B T}{2\pi} \right)^{\frac{3}{2}},
\]
i.e. the bosonic capacity \(\sim T^{\frac{3}{2}}\).
On the other hand, if the particles are fermions, then the capacity, for \(T \leq T_f\) (where in this case, 
\(T_f= \frac{\hbar^2}{2mk_B} \left( \frac{6\pi^2N}{gV} \right)^{\frac{2}{3}}\)), is given by (see e.g. \cite{molla-nasiruddin})
\[
\frac{C^{fd}_{V,N,T}}{\log_2 e} = \frac{\pi^2}{2} \frac{T}{T_f},
\]
i.e. the fermionic capacity scales as \(T\).
Clearly, the fermionic capacity is higher than the bosonic one for sufficiently low temperatures in a bounded volume, such as a 3D box.

In the case of a harmonic trap in 3D, the bosonic capacity scales as \(T^3\) (see e.g. \cite{Stringari-r_songe_kothha_holo}),
when the temperature is below critical.
For 
sufficiently low temperatures, the fermionic capacity scales as \(T\), which can be estimated by 
using the Sommerfeld expansions of Fermi functions.
Therefore the fermionic capacity wins once again.

\section{Conclusions}
\label{sec-sesh}

In this paper, we have investigated the classical as well as the quantum capacity of noiseless quantum channels, carrying massive
  particles. 
We have considered spinless noninteracting 
bosons and fermions. Noninteracting bosons exhibit Bose Einstein condensation, and we have shown that this critical behavior can also
be observed by studying the capacity  of a quantum channel carrying bosons. We show that the capacity of such channels 
is concave (with respect to temperature) above the critical temperature, while it is convex (with respect to temperature) below that point.
This criticality is absent in the case of fermions, as expected. 
We have numerically evaluated the capacities of bosons and fermions for different numbers of particles. 
In the case of bosons, even in the case of a small number of particles, say \(100\), condensation can be observed from the qualitative change 
in behavior of the capacity.

We have also  shown analytically that a channel carrying bosons  is not as good a medium for transferring
classical as well as quantum information, as a channel carrying fermions. This is true for a wide range of potentials that can be 
currently created in the laboratory. 
The analytical calculation for power law potentials holds for  moderate temperatures. 
However, numerically we have checked that this is true even for low temperatures. 
It is tempting to believe that  
such superiority of fermions over bosons is generic, at least for  power-law potentials. 
In special cases, we have considered the low temperature behavior analytically, and have shown that for sufficiently low temperatures, the fermionic capacity is 
higher than the bosonic one, for the 3D box and the 3D harmonic trap.


\begin{acknowledgments}

We thank Sandro Stringari for important discussions, and the Horodeccy Family for sending us a 
 draft of Ref. \cite{Horodecki_private}.
We acknowledge support from the
Deutsche Forschungsgemeinschaft 
(SFB 407, SPP 1078, SPP 1116, 436POL), 
the Alexander von Humboldt Foundation, the Spanish MEC grant FIS-2005-04627,
the ESF Program QUDEDIS, and EU IP SCALA.
This work was partially supported by the Polish Ministry of Scientific
Research and Information Technology under Grant No. PBZ/MIN/008/P03/2003 and
by the University of Lodz.

\end{acknowledgments}


\begin{thebibliography}{99}
\bibitem{bayonet-hok-joto-dharalo-kaste-ta-shan-diyo-bondhu} This is because, the last relation that 
we use to obtain the relation (\ref{robimama-dei-hama-gaye-ranga-jama-oi}), is 
\(\log_2 \Omega((n-p+1)E, (n-p+1)N) = \log_2 \Omega ((n-p)E, (n-p)N) + S(E,N)\), which assumes that 
\(n-p\) is large.
\bibitem{Shell-aar-bomb-hok-joto-bharalo-kaste-ta-shan-diyo-bondhu} For example, one may choose \(p=t - {\cal b} \log_e t {\cal c}\), and 
\(n=t\), for any positive integer \(t\), where \({\cal b} \cdot {\cal c}\) denotes the integer 
part of its argument.
\bibitem{aaj-bikeley-ki-bhai-h(n)at_tey-jabi-na-jabina}
To arrive at Eq. (\ref{Niren-Chattopadhyay}) from 
Eq. (\ref{Srikant}), 
we transform the cartesian momentum coordinates  in the integral in Eq.
(\ref{Srikant})
to hyperspherical coordinates, to have 
\begin{eqnarray}
\varsigma(\epsilon) &=& 
\mbox{constant} \cdot  \int p^{d-1}dp d^d x \phantom{a} \delta \left(\frac{p^2}{2m} + V(\vec{r}) - \epsilon \right) \nonumber \\
&=& \mbox{constant} \cdot  \int \varpi^{(d-2)/2} d \varpi d^d x \phantom{a} \delta \left(\varpi + V(\vec{r}) - \epsilon \right), \nonumber 
\end{eqnarray}
where we have substituted \(p^2/2m = \varpi\). Eq. (\ref{Niren-Chattopadhyay}) follows directly from the last equation.
\end{thebibliography}
\end{document}